\documentclass[twocolumn,pra,  superscriptaddress]{revtex4-1}
\usepackage{graphicx}
\usepackage{braket}
\usepackage{amsmath,amsfonts,amssymb,amstext,amscd,amsthm,bbold}
\usepackage{comment,float}

\usepackage[colorlinks=true,linkcolor=blue,urlcolor=magenta,citecolor=magenta]{hyperref}
\usepackage[normalem]{ulem}

\begin{document}
\title{Digital quantum simulation framework for energy transport in an\\ open quantum system}
\author{Pragati Gupta}
\email{pragatigupta@iisc.ac.in}
\affiliation{Indian Institute of Science, C.V. Raman Avenue, Bengaluru 560012, India}
\author{C. M. Chandrashekar}
\email{chandru@imsc.res.in}
\affiliation{The Institute of Mathematical Sciences, C. I. T. Campus, Taramani, Chennai 600113, India}
\affiliation{Homi Bhabha National Institute, Training School Complex, Anushakti Nagar, Mumbai 400094, India}
\affiliation{Dept. of Instrumentation \& Applied Physics, Indian Institute of Sciences, C.V. Raman Avenue, Bengaluru 560012, India}
\begin{abstract}
Quantum effects such as the environment  assisted  quantum  transport  (ENAQT) displayed in photosynthetic Fenna-Mathews-Olson (FMO) complex 
 has  been  simulated  on  analog  quantum simulators.  Digital quantum simulations offer greater universality and flexibility over analog simulations.  However, digital quantum simulations of open quantum systems face a theoretical challenge; one does not know the solutions of the continuous time master equation for developing quantum gate operators.  We give a theoretical framework for digital quantum simulation of ENAQT by introducing new quantum evolution operators.  We develop the dynamical equation for the operators and prove that it is an analytical solution of the master equation.  As an example, using the dynamical equations, we simulate the FMO complex in the digital setting, reproducing theoretical and experimental evidence of the dynamics.  The framework gives an optimal method for {quantum circuit} implementation, giving a log reduction in complexity over known methods.  The generic framework can be extrapolated to study other open quantum systems.
\end{abstract}
\maketitle

\section{\label{chap1}Introduction}
\noindent
Quantum simulators are devices that can turn the exponential scaling of resources needed to simulate and understand properties of complex quantum systems on classical computers into a favourable polynomial overhead.  Original proposal for quantum computers by Feynman\,\cite{feynman} was to simulate complex quantum systems such as many body quantum systems in low temperature physics and lattice gauge theories.  An algorithm by Shor to solve the discrete logarithm problem on a quantum computer\,\cite{shor} highlighted the broader application of quantum computers. However, a full scale quantum computer with millions of qubits is far from being realised\,\cite{preskill}. Ongoing experimental research is geared towards making quantum simulators with small number of qubits\,\cite{smith} as first practical applications of a quantum computer. Quantum simulators can be analog or digital. Analog simulations use continuous time evolution models such as the Lindblad master equation\,\cite{digitalanalog}. Digital simulations are done using quantum gates for simulating discrete time evolution operators. Compared to analog quantum simulators, digital quantum simulations offer universality and flexibility\,\cite{digitalanalog}. One of the major challenges in building a quantum simulator has been the presence of noise due to environment. However, recently a number of studies have remarkably shown that noise can facilitate transmission of energy in quantum simulators through a process called environment assisted quantum transport (ENAQT)\,\cite{ENAQT}. Environment engineering in quantum networks for enhanced energy transport has been studied to understand the role of noise in ENAQT\,\cite{uchiyama2018environment} and for building quantum simulators\,\cite{NAT}.

One of the important application of quantum simulators in near-term is to understand the dynamics in chemical systems. Chemical complexes have been found to show a variety of quantum effects\,\cite{quantumchem} such as ENAQT. The Fenna-Mathews-Olson (FMO) protein complex is one of the most widely studied photosynthetic systems. It transfers the excitation energy from light harvesting antenna to the reaction centre in the photosynthetic complex of certain bacteria\,\cite{visualization}.  FMO complex has been studied for decades to understand the underlying mechanism of high efficiency excitation energy transfer (EET)\,\cite{efficiency} in photosynthetic complexes\,\cite{theoryfmo, engel, forster, redfield1, redfield2, redfield3, modified, stochastic, comparison}. The phenomenon of delocalized exciton states has been observed via electronic spectroscopy in light-harvesting complexes\,\cite{theoryfmo} and the FMO protein complex\,\cite{cho}. A combination of quantum and {dissipative} effects have been observed to felicitate the transfer of excitation energy through the complex\,\cite{cho}. Engel et al.\,\cite{engel} found direct evidence for quantum coherence in the FMO complex,  Mohseni et al.\,\cite{mohseini} gave a continuous-time quantum walk model with environment assisted transport (ENAQT) and Plenio et al.\,\cite{PH08} presented a dephasing-assisted transport model to successfully explain the high efficiency. While quantum walks provide models for quantum simulation\,\cite{aharanov}, continuous-time quantum walk over noisy lattices has been studied for understanding spatially correlated noise\,\cite{rossi2017continuous}. Quantum simulations of FMO complex have been done to understand simulations of open quantum systems\,\cite{Childs} and other systems in quantum chemistry\,\cite{kosenkov}. It is also being studied to develop efficient artificial light-harvesting systems\,\cite{baker}.  Analog quantum simulations of FMO complex have been done on NMR quantum computer\,\cite{Wang_2018}, superconducting qubits\,\cite{Mostame}, superconducting circuits\,\cite{Anton} and ultracold atoms\,\cite{ross}. In \,\cite{mahdian,mahdian2}, Mahdian et al. described a setup for digitally simulating the FMO complex. But their study does not include quantum jumps and the interplay of quantum and dissipative effects which are the salient features of the environment assisted quantum walk. These previous studies led us to explore and characterize the theoretical model with a general framework for digital quantum simulation of the dynamics in chemical complexes along with FMO complex as a specific example. 

Open quantum systems\,\cite{theoryopensystems} are often described using a reduced density matrix which is obtained by tracing out the environment from density matrix of the whole system in a closed form (system + environment). The master equation is used to describe the evolution of the reduced density matrix. Analog simulations of open quantum systems directly mimic the continuous-time evolution of the quantum system of interest on the simulator\,\cite{Mostame}. Master equations are commonly used models for analog simulations. Other approaches like the Quantum Langevin equations\,\cite{langevin1, langevin2} based on the Heisenberg approach give operator equations for describing open quantum systems. Since simulations depend on evolving the simulator to mimic the system of interest, Langevin equations need to be remodelled to equivalent master equation for quantum simulation. Thus, master equations serve as a more suitable model for quantum simulation of open quantum systems. This approach is more suitable for modelling quadratic Hamiltonians over modelling of non-quadratic interacting Hamiltonians for which further probe is needed. Here, we will focus on quadratic Hamitonians. Numerical simulation of open quantum systems can be done using the QuTIP\,\cite{johansson2012qutip} package for python, which uses master equation or other continuous-time models. Digital quantum simulation of open quantum system can be achieved by using an operator sum representation of the dynamics\,\cite{1}. The operator sum representation is much more general than master equations or other models. For example, non trace preserving processes can be simulated by adding an extra dummy operator to complete the trace. Such techniques are not possible for master equations. The operator sum representation can also simulate non Markovian dynamics, unlike Lindblad master equation. Thus, apart from being suitable for digital quantum simulation, the operator sum representation offers several advantages for numerical simulation as they can capture a wider range of phenomena. 

In this work, we develop a theoretical framework for digital simulation of environment assisted energy transfer in open quantum systems. The main challenge in developing discrete time evolution operators for digital quantum simulation is that one needs to solve the master equation for which solutions are otherwise not known. We give a methodology to develop evolution operators and dynamical equation in the operator sum representation which capture the interplay of unitary quantum evolution and noise in open quantum systems. We derive the discrete-time evolution operators and dynamical equation for the process of ENAQT using this methodology and give a mathematical proof that the derived dynamical equation is the analytical solution of the Lindblad master equation. The evolution equation is generalised to incorporate variable strength of system-bath interaction which helps in controlling the dynamics by tuning the level of noise. This gives a theoretical model for the digital quantum simulation of ENAQT with tunable bath coupling. As an application of this framework we give a discrete-time dynamical model for the FMO complex. Energy transfer in the FMO complex happens through environment assisted quantum transport of delocalized excitons over a network of strongly coupled sites and is dependent on the temperature of environment. We give the dynamical equation to capture the energy transfer and demonstrate its high efficiency through numerical simulations. We use the tunable bath-coupling model to simulate temperature dependence of the dynamics and show that the results are consistent with existing theoretical and experimental evidence. Finally, we give the quantum circuit for the implementation of the dynamical equation and discuss it's space and time complexity. We show that our framework gives a log-space reduction in complexity over existing techniques.

This article is organised as follows. The basic description of ENAQT is given in Sec.\,\ref{Chap2.secENAQT}. In Sec.\,\ref{Chap3} theoretical framework for digital quantum simulation of ENAQT is developed. In Sec.\,\ref{Chap4}, the framework is applied to model ENAQT in the FMO complex and  simulations of this model are presented. Some concluding remarks are given in Sec.\,\ref{sec5}.

\section{Environment assisted quantum transport}\label{Chap2.secENAQT}

Hamiltonians describing the dynamics of the quantum systems typically possess energy mismatches, that can hinder transmission of excitation due to Anderson localization\,\cite{anderson}. However, quantum systems are also generally subjected to relatively high levels of environment-induced noise and decoherence. A certain degree of noise can cause transfer of excitation through {dissipative} processes, which can overcome localization. The interplay between the coherent dynamics of the system and the incoherent action of the environment can result in greater transport efficiency than coherent dynamics on its own\,\cite{interplay}. The noise is in the form of relaxation and dephasing, combined with coherent dynamics leads to the phenomenon of ENAQT.  
The Hamiltonian of the system due to pure quantum interactions is given by,
\begin{equation}
    H =\sum_{m} \varepsilon_m \ket{m}\bra{m} + \sum_{n<m} V_{mn}\Big(\ket{m}\bra{n} + \ket{n}\bra{m}\Big)
\end{equation}
where $\varepsilon_m$ are the energies of the states and $V_{mn}$ denotes the coupling leading to coherence between different states.
The main effects of the environment, dephasing and relaxation together lead to quantum jumps between states. Quantum jumps are represented by the following phonon bath Hamiltonian,
\begin{equation}
    H_p=\sum_{m,n} q_{mn}\ket{m}\bra{n}
\end{equation}
where $q_{mn}$ are the couplings due to the phonon bath. The mathematical form of both, $V_{mn}$ and $q^p_{mn}$ are same, however $V_{mn}$ gives the coherent couplings between the sites, while $q^p_{mn}$ gives the rate of quantum jumps between sites due to the environment interaction. $V_{mn}$ only depends on the system of chromophores and leads to coherent evolution of the system between different sites. The $q^p_{mn}$ depends on couplings of the system to the environment and is used to obtain probabilities of quantum jumps. \\
This leads to the Lindblad master equation,
\begin{equation}\label{eq:masterENAQT}
    \begin{split}
    \frac{d\rho}{dt}=\mathcal{L}(\rho)=-i&[H,\rho]+L(\rho),\\
    L(\rho)=\sum_{m,n}\gamma_{mn}\Big(L_{mn}\rho L_{mn}^\dagger
    &-\frac{1}{2}L_{mn}L_{mn}^\dagger\rho - \frac{1}{2}\rho L_{mn}L_{mn}^\dagger\Big)
    \end{split}
\end{equation}
where $L_{mn}=\ket{m}\bra{n}$ and $\gamma_{mn}$ is obtained from $q_{mn}$.
\vskip 0.2cm
The dynamics of ENAQT depend on quantum coherence, described by the system Hamiltonian and on the environment induced quantum jumps between different states. The environment causes decoherence of the states. These effects together cause a transfer of excitation energy in ENAQT. The overall dynamics due to environment are non unitary and trace preserving. 
\section{\label{Chap3}Framework for environment assisted quantum transport} 
\subsection{Open quantum systems}\label{Chap2.opensystem}
We can represent a system in interaction with the environment as an open quantum system, where the  environment is modelled by a phonon bath. An open quantum system is a part of a larger closed system in Hilbert space $\mathcal{H}= \mathcal{H}_S\otimes \mathcal{H}_B$, where $\mathcal{H}_B$ is the phonon bath Hilbert space\,\cite{2}. Assuming the initial state is represented by the separable density matrix $\rho = \rho_S \otimes|0\rangle\langle0|_B$, the evolution of the total system is,
\begin{equation*}\label{eq:opensystem}
    \rho(t) = U_{SB} (\rho_S \otimes|0\rangle\langle0|_B )U_{SB}^\dagger.
\end{equation*}
A partial trace over B gives the evolution of the open quantum system S, 
\begin{equation*}\label{eq:trace}
\begin{split}
    \rho_S (t) = Tr_B (\rho(t))= \sum_k \langle k|U_{SB}(\rho_S \otimes|0 \rangle \langle 0|_B )U_{SB}^\dagger |k\rangle \\
    = \sum_k \langle k|U_{SB}|0\rangle\rho_S (0)\langle 0|U_ {SB}^\dagger |k\rangle
\end{split}
\end{equation*}
which in the form of Kraus operators $M_k$ will be, 
\begin{equation}\label{eq:Krausequation}
    \rho_S (t) = M(\rho_S(0)) = \sum_k M_k\rho_S (0)M_k^\dagger
\end{equation}
where
\begin{equation}\label{eq:Kraus}
    M_k=\langle k|U_{SB} |0\rangle = Tr_B (|0\rangle \langle k|U_{SB}).
\end{equation}
Here $|k\rangle $  is an orthonormal basis for $\mathcal{H}_B$ and $\sum_kM_k^\dagger M_k=\mathbb{1}$. This can be used to obtain operators for effects of environment on the system by introducing a bath to describe the dynamics and then tracing it out. However, ENAQT is a combination of quantum and {dissipative} effects, described by the Lindblad master equation, Eq.\,(\ref{eq:masterENAQT}). We need to solve the master equation to obtain operators for the combined dynamics.

\subsection{\label{Chap3.analytical}Analytical solution} 
To solve master equation, one needs to obtain operators which capture the combined effect of quantum and {dissipative} processes. This can be obtained by combining the operators for different processes of ENAQT. We develop the following methodology to systematically derive the operators and the dynamical equation for ENAQT.\\
\\
{\it{Methodology:}\label{methodology}} To model both quantum and {dissipative} effects we introduce evolution operators which are a combination of Kraus operators and unitary quantum evolution. Using these operators, we derive the evolution equation for density matrix. We develop the general model by first taking a toy system and step by step adding different processes to it, to arrive at the final picture. We consider a system with one quantum jump and find a model for this toy system using the following procedure: We write an evolution for system + bath which appropriately captures the quantum jump. We trace out the bath from the evolution and find the Kraus operators for quantum jump in the system. Then, we introduce  unitary quantum evolution to this system in addition to the quantum jump. Thus, the new evolution operators are obtained by appropriately combining Kraus operators and unitary quantum evolution operator. We use these operators to write the discrete time evolution equation for this setup, which appropriately captures the interplay of quantum and {dissipative} effects.

Next, we add another quantum jump to the system to generalise the toy model to simulate multiple quantum jumps. Again, we draw a parallel from the first case and follow the above procedure to develop a model for this setup. We write the evolution operators and arrive at the dynamical equation for this setup. This gives a general model for simulating ENAQT in open quantum systems with unitary quantum evolution and multiple environment induced quantum jumps. 
\subsubsection{Toy Model: Single quantum jump}\label{Chap3.toy}
Consider a two level system with states given by $|0\rangle_S$ and $|1\rangle_S$. Say the bath B induces a quantum jump from $|0\rangle_S$ to $|1\rangle_S$ with probability $p_{0\to 1}$. The system + bath evolution can be formalised as follows,
\begin{equation*}\label{eq:case1}
\begin{split}
    |0\rangle _S|0\rangle _B & \to \sqrt{1-p_{0\to 1}}|0\rangle _S|0\rangle _B +\sqrt{p_{0\to 1}}|1\rangle _S|1\rangle _B\\
    |1\rangle _S|0\rangle _B & \to |1\rangle _S|0\rangle _B .
\end{split}
\end{equation*}
The Kraus operators for quantum jumps on the system, obtained from tracing out the bath are,\\

$\qquad M_0$=$
\begin{pmatrix}
\sqrt{1-p_{0\to 1}} & 0 \\
 0 & 1
\end{pmatrix}~~~$  and    $~~~M_1$=$
\begin{pmatrix}
0 & 0 \\
\sqrt{p_{0\to 1}} & 0
\end{pmatrix}$.\\
The operators can also be written as,
\begin{equation}\label{eq:case1kraus}
\begin{split}
    &M_0=\sqrt{1-p_{0\to 1}}|0\rangle \langle 0| + |1\rangle \langle 1|  ~~~~\mbox{and}\\
    ~~~~ &M_1=\sqrt{p_{0\to 1}}|1\rangle \langle 0|.
\end{split}
\end{equation}
Now, introduce quantum evolution to the system. The Lindblad equation, with the system being subject to free Hamiltonian $H_S$ is,
\begin{equation}\label{eq:case1mastereq}
\frac{\partial\rho(t)}{\partial t} = -\frac{i}{\hbar}[H_S,\rho(t)] + L(\rho(t)),
\end{equation}
$L(\rho(t))$ is due to the effect of quantum jumps, where
\begin{equation}\label{eq:case1lindblad}
    L(\rho(t))= \sum_k[ L^k\rho L^{k\dagger}-\frac{1}{2}L^kL^{k\dagger}\rho - \frac{1}{2}\rho L^kL^{k\dagger}].
\end{equation}
Here $L^k$ are quantum jump operators. The terms carry their usual meaning, first term represents the quantum jumps and the other two terms are normalisation terms for the case when the jump does not happen. The Kraus operators can be combined with unitary quantum evolution as follows to solve this master equation,
\begin{equation}\label{eq:case1operators}
    M_0'= M_0U \quad\quad M_1'=M_1
\end{equation}
where {$U= e^{-\frac{iH_S\Delta t}{\hbar}}$}. It can be verified that $M_0'^\dagger M_0' + M_1'^\dagger M_1' =\mathbb{1}$. These evolution operators,  Eq.(\ref{eq:case1operators}), where $M_0'$ has unitary quantum evolution and $M_1'$ does not, will be useful to capture the appropriate dynamics, as will be proved in the following section. The operators can be interpreted as follows. Two processes are happening in the system at any time t: coherent evolution and quantum jumps. From a state, the population density can move out of the state via these two processes. From the total population, some goes out via quantum jumps. And from the population remaining after the quantum jumps, some goes out via quantum coherence. The second operator $M_1'$ depicts the moving out through quantum jumps. $M_0$ in the first operator capture's the population that remains after the quantum jump. From this remaining population, some moves out via quantum coherence as captured by $U$ in $M_0'=M_0U$.\\
Thus, the discrete time density matrix evolution equation obtained from these operators is given by,
\begin{equation}\label{eq:case1operatoreq}
\begin{split}
    \rho(t+\Delta t) &= M_0'(\sqrt{\Delta t})\rho(t)M_0'^{\dagger}(\sqrt{\Delta t}) \\
    &+ M_1'(\sqrt{\Delta t})\rho(t)M_1'^{\dagger}(\sqrt{\Delta t}).
\end{split}
\end{equation}
Here $\sqrt{\Delta t}$ is taken, so that considering contributions from $M_0'$ and $M_0'^\dagger$ and assuming linear dependence on time, the net change is of the first order in $\Delta t$.

Substituting Eq.(\ref{eq:case1kraus}) in Eq.(\ref{eq:case1operatoreq}),
\begin{equation}\label{eq:case1eq}
\begin{split}
    \rho(t+\Delta t)=(1-p_{0\to 1})|0\rangle \langle 0|U\rho(t) U^\dagger|0\rangle \langle 0|&\\
    +\sqrt{1-p_{0\to 1}}|0\rangle \langle 0|U\rho(t) U^\dagger|1\rangle \langle 1|&\\
    +\sqrt{1-p_{0\to 1}}|1\rangle \langle 1|U\rho(t) U^\dagger|0\rangle \langle 0|&\\
    +|1\rangle \langle 1|U\rho(t) U^\dagger|1\rangle \langle 1|&\\
    +(p_{0\to 1})|1\rangle \langle 0|\rho(t)|0\rangle \langle 1|&.
\end{split}
\end{equation}
The second and the third terms of the preceding equation capture the decoherence due to the bath and contribute to the normalisation terms. The first term and the last term are due to the quantum jump. Correspondence of the different terms to the Lindblad equation given above can be seen here. $U$ and $U^\dagger$ capture the quantum coherence.

\subsubsection{General Model: Multiple quantum jumps }\label{Chap3.generalmodel}
We can generalise the previous setup by including a quantum jump in the other direction, from $|1\rangle_S$ to $|0\rangle_S$ with probability $p_{1\to 0}$. This can be represented using a two qubit bath, where the the first qubit ($B_1$) captures the first quantum jump and qubit, $B_2$ captures the second quantum jump. Then the system + bath evolution is given by,
\begin{equation}\label{eq:case2}
\begin{split}
    |0\rangle _S|0\rangle _{B_1}|0\rangle _{B_2} \to  \sqrt{1-p_{0\to 1}}|0\rangle _S|0\rangle _{B_1}|0\rangle _{B_2}&\\
    +\sqrt{p_{0\to 1}}|1\rangle _S|1\rangle _{B_1}|0\rangle _{B_2}& ~;~\\
    |1\rangle _S|0\rangle _{B_1}|0\rangle _{B_2} \to  \sqrt{1-p_{1\to 0}}|1\rangle _S|0\rangle _{B_1}|0\rangle _{B_2}& \\
    +\sqrt{p_{1\to 0}}|0\rangle _S|0\rangle _{B_1}|1\rangle _{B_2}&
\end{split}
\end{equation}
and the corresponding Kraus operators for the system, after tracing out the bath are,
\begin{equation}\label{eq:case2kraus}
\begin{split}
    M_{00}=\sqrt{1-p_{0\to 1}}|0\rangle \langle 0|    &\qquad  M_{01}=\sqrt{p_{0\to 1}}|1\rangle \langle 0|\\
    M_{11}=\sqrt{1-p_{1\to 0}}|1\rangle \langle 1| & \qquad M_{10}=\sqrt{p_{1\to 0}}|0\rangle \langle 1|.
\end{split}
\end{equation}
Adding quantum evolution to Kraus operators when the system is also subject to free Hamiltonian $H_S$, similar to the previous case leads to the following evolution operators,
\begin{equation}\label{case2operators}
\begin{split}
    M_{00}'=M_{00}U\qquad &M_{01}'=M_{01} \\
    M_{11}'=M_{11}U\qquad  &M_{10}'=M_{10}.
\end{split}
\end{equation}
Drawing parallel from Eq.(\ref{eq:case1eq}), the discrete time dynamical equation is given by,
\begin{equation}\label{eq:case2eq}
\begin{split}
    \rho(t+\Delta t)=(1-p_{0\to 1})|0\rangle \langle 0|U\rho(t) U^\dagger|0\rangle \langle 0|&\\
    +\sqrt{1-p_{0\to 1}}\sqrt{1-p_{1\to 0}}|0\rangle \langle 0|U\rho(t) U^\dagger|1\rangle \langle 1|&\\
    +\sqrt{1-p_{0\to 1}}\sqrt{1-p_{1\to 0}}|1\rangle \langle 1|U\rho(t) U^\dagger|0\rangle \langle 0|&\\
    +(1-p_{1\to 0})|1\rangle \langle 1|U\rho(t) U^\dagger|1\rangle \langle 1|&\\
    +(p_{0\to 1})|1\rangle \langle 0|\rho(t)|0\rangle \langle 1|&\\
    +(p_{1\to 0})|0\rangle \langle 1|\rho(t)|1\rangle \langle 0|&.
\end{split}
\end{equation}
The last two terms, and the first and the fourth terms above represent the quantum jumps. The second and third terms capture the decoherence and contribute to the normalisation terms.  $U$ and $U^\dagger$ capture the unitary quantum evolution.\\
Substituting Eq.(\ref{case2operators}) in Eq.(\ref{eq:case2eq}), we arrive at the general model for simulating quantum jumps and unitary quantum evolution,\\
\begin{equation}\label{eq:case2operatoreq}
\begin{split}
    \rho(t+\Delta t) = &M_{00}'(\sqrt{\Delta t})\rho(t)M_{00}'^{\dagger}(\sqrt{\Delta t}) \\
    +&M_{00}'(\sqrt{\Delta t})\rho(t)M_{11}'^{\dagger}(\sqrt{\Delta t})\\
    +&M_{11}'(\sqrt{\Delta t})\rho(t)M_{00}'^{\dagger}(\sqrt{\Delta t})\\
    +&M_{11}'(\sqrt{\Delta t})\rho(t)M_{11}'^{\dagger}(\sqrt{\Delta t})\\
    +&M_{01}'(\sqrt{\Delta t})\rho(t)M_{01}'^{\dagger}(\sqrt{\Delta t})\\
    +&M_{10}'(\sqrt{\Delta t})\rho(t)M_{10}'^{\dagger}(\sqrt{\Delta t}).
\end{split}
\end{equation}
The second and third terms above are the crucial features of this dynamical equation. These asymmetric terms with $M_{00}'$ and $M_{11}'$ representing decoherence make this equation conceptually different from the evolution equation obtained from Kraus operators,
\begin{equation*}
    \rho_S (t) = \sum_k M_k\rho_S (0)M_k^\dagger.
\end{equation*}
This evolution equation does not have asymmetric terms of the form $M_i\rho_S(0)M_j^\dagger$, which are present in the equation we have arrived at, Eq.\eqref{eq:case2operatoreq} by drawing a parallel, Eq.(\ref{eq:case2eq}) from the case with single quantum jumps, Eq.(\ref{eq:case1eq}). Now we can see that this parallelism gave us insight to add these asymmetric terms to Eq.(\ref{case2operators}), which would have been hard to see directly from the evolution equation for Kraus operators, Eq.(\ref{eq:Krausequation}). This shows that the methodology used in developing the model is effective to flesh out the finer details of the theoretical framework. Eq.\eqref{case2operators} can be applied to digitally simulate any open quantum system with environment assisted evolution.  It is the discrete time solution of the Lindblad master equation.
\subsection{\label{Chap3.proof}Proof of correctness} 
The equivalence of the dynamical equations,  Eq.(\ref{eq:case1operatoreq}) and (\ref{case2operators}) to the Lindblad formalism, Eq.(\ref{eq:case1mastereq}) is proven in this section. Considering the case with single quantum jump, the following derivation shows that Eq.(\ref{eq:case1operatoreq}) is the discrete time solution to Lindblad master equation governing the system. 

The master equation is in continuous time formalism. For continuous time limit of the discrete dynamics substituting $\Delta t \to \partial t$ in Eq. (\ref{eq:case1operatoreq}),
\begin{equation}\label{eq:case1sol}
\begin{split}
    \rho(t+\partial t) = &M_0'(\sqrt{\partial t})\rho(t)M_0'^{\dagger}(\sqrt{\partial t}) \\
    + &M_1'(\sqrt{\partial t})\rho(t)M_1'^{\dagger}(\sqrt{\partial t}).
\end{split}
\end{equation}
Using Eq.(\ref{eq:case1operators}),
\begin{equation}\label{eq:proofeq1}
\begin{split}
    M_0'(\sqrt{\partial t})=M_0U=M_0(\sqrt{\partial t}) e^{-\frac{iH\partial t}{\hbar}}&\\
    \approx M_0(\sqrt{\partial t})\Big [\mathbb{1}-\frac{iH\partial t}{\hbar}\Big]&.
\end{split}
\end{equation}
Now, $M_0M_0^{\dagger} + M_1M_1^{\dagger}=\mathbb{1} $ and $M_0 = M_0^\dagger$, so $M_0(\sqrt{\partial t})$ can be written as,
\begin{equation}\label{eq:proofeq2}
\begin{split}
    M_0(\sqrt{\partial t})\approx \sqrt{\mathbb{1}- M_1(\sqrt{\partial t})M_1^{\dagger}(\sqrt{\partial t}})&\\
    \approx \mathbb{1}-\frac{1}{2} M_1(\sqrt{\partial t})M_1^{\dagger}(\sqrt{\partial t})&.
\end{split}
\end{equation}
Substituting Eq.(\ref{eq:proofeq2}) in Eq.(\ref{eq:proofeq1}),
\begin{equation}\label{eq:proofeq3}
\begin{split}
    M_0'&(\sqrt{\partial t})\approx M_0(\sqrt{\partial t})\Big[\mathbb{1}-\frac{iH\partial t}{\hbar}\Big] \\
    &\approx \Big[\mathbb{1}-\frac{1}{2} M_1(\sqrt{\partial t})M_1^{\dagger}(\sqrt{\partial t})\Big]\Big[\mathbb{1}-\frac{iH\partial t}{\hbar}\Big].
\end{split}
\end{equation}
Considering terms only up to first order in $\partial t$, we get,
\begin{equation}\label{eq:proofeq4}
\begin{split}
    M_0'(\sqrt{\partial t})&\approx \Big[\mathbb{1}-\frac{iH\partial t}{\hbar}-\frac{1}{2} M_1(\sqrt{\partial t})M_1^{\dagger}(\sqrt{\partial t})\Big]\\
    M_0'^\dagger(\sqrt{\partial t})&\approx \Big[\mathbb{1}+\frac{iH\partial t}{\hbar}-\frac{1}{2} M_1(\sqrt{\partial t})M_1^{\dagger}(\sqrt{\partial t})\Big].
\end{split}
\end{equation} 
Substituting $M_0'$, Eq.(\ref{eq:proofeq4}) and $M_1'$, Eq.(\ref{eq:case1operators}) in Eq.(\ref{eq:case1sol}),
\begin{equation}\label{eq:proofeq5}
\begin{split}
    \rho(t+\partial t)\approx \Big [\Big(\mathbb{1}-\frac{iH\partial t}{\hbar}-\frac{1}{2} M_1(\sqrt{\partial t})M_1^{\dagger}(\sqrt{\partial t})\Big)*&\\
    \rho(t)\Big(\mathbb{1}+\frac{iH\partial t}{\hbar}-\frac{1}{2} M_1(\sqrt{\partial t})M_1^{\dagger}(\sqrt{\partial t})\Big)\Big]&\\
    +M_1(\sqrt{\partial t})\rho(t)M_1^{\dagger}(\sqrt{\partial t})&.
\end{split} 
\end{equation}
Considering terms only up to first order in $\partial t$ and rearranging the terms,
\begin{equation}\label{eq:proofeq6}
\begin{split}
    \rho(t+\partial t) = \rho(t) -\frac{i}{\hbar}\partial t[H,\rho (t)]\qquad\qquad& \\
    +M_1(\sqrt{\partial t})\rho(t)M_1^{\dagger}(\sqrt{\partial t})&\\
    -\frac{1}{2} M_1(\sqrt{\partial t})M_1^{\dagger}(\sqrt{\partial t})\rho(t)&\\
    -\frac{1}{2}\rho(t)M_1(\sqrt{\partial t})M_1^{\dagger}(\sqrt{\partial t})&.
\end{split}
\end{equation}
We can see that the evolution operators,  Eq.(\ref{eq:case1operators}) helped us get the quantum jump and the normalisation terms correctly. Padding $M_1'$ as well with unitary quantum evolution($U$) would lead to extra unwanted terms.\\
\\
Set $ M_1(\sqrt{\partial t})=L_1 \sqrt{\partial t}$,
\begin{equation}\label{eq:proofeq7}
\begin{split}
    \rho(t+\partial t) = \rho(t) +\partial t\Big[-\frac{i}{\hbar}[H,\rho (t)]+L_1\rho(t)L_1^{\dagger}& \\
    -\frac{1}{2} L_1L_1^{\dagger}\rho(t)- \frac{1}{2}\rho(t)  L_1L_1^{\dagger}\Big]&.
\end{split}
\end{equation}
We obtain the Lindblad equation,
\begin{equation*}\label{eq:proofeq8}
\begin{split}
    \frac{\partial\rho(t)}{\partial t} &= -\frac{i}{\hbar}[H_S,\rho(t)] + L(\rho(t))\\
    \Big(L(\rho(t))&=L_1\rho(t)L_1^{\dagger} -\frac{1}{2} L_1L_1^{\dagger}\rho(t) - \frac{1}{2} \rho(t)L_1L_1^{\dagger}\Big).
\end{split}
\end{equation*}
We started from the dynamical equation, Eq.(\ref{eq:case1operatoreq}) and arrived at the Lindblad master equation, Eq.(\ref{eq:case1mastereq}). Thus, for  the case with single quantum jump, we showed that the dynamical model is the solution for its Lindblad equation in the Markov approximation. A similar proof for the more general case with multiple quantum jumps, Eq.(\ref{eq:case2operatoreq}) can be worked out. \\

Thus, Eqs.(\ref{case2operators}) and (\ref{eq:case2operatoreq}) can be used to describe the complete dynamics of environment assisted quantum walk in general open quantum systems. 
%
%
%
%
\subsection{Model for tunable bath coupling \label{Chap4.tunable}}
\noindent
The dynamics of ENAQT depend on the level of noise the system is subject to. Noise depends on the strength with which the system couples to the bath. At lower level of coupling, the strength is a fraction of the full coupling mode. We can denote this fraction as $\chi\in [0,1]$. The coupling strength dependent master equation can be written as,
\begin{equation}\label{eq:master_chi}
    \frac{\partial \rho(t)}{\partial t} = -\frac{i}{\hbar}[H_c,\rho(t)] + \chi L_p(\rho(t)).\end{equation}
This can be modified to be written as,
\begin{equation}\label{eq:master_chi2}
    =-(1-\chi)\frac{i}{\hbar}[H_c,\rho(t)] + \chi \Big[-\frac{i}{\hbar}[H_c,\rho(t)]+L_p(\rho(t)\Big].
\end{equation}
For $\chi=1$ this reduces to the normal master equation. For other values of $\chi$, it captures the dynamics at different strengths of system-bath coupling. This can be digitally simulated by the following dynamical equation,
\begin{equation}\label{eq:fmo_chi}
    \rho_\chi(t +\Delta t)=(1-\chi)U\rho(t) U^\dagger
    +\chi\rho(t +\Delta t)
\end{equation}
where $\rho(t +\Delta t)$ is given by Eq.(\ref{eq:case2operatoreq}). $\chi$ can be used to study variation of the dynamics with respect to changes in the system-bath interaction. Thus we obtain the discrete time dynamical equation for digital quantum simulation of environment assisted quantum walk with variable bath coupling. In the next section, we apply this model to simulate the FMO complex.\bigskip
%
%
%
%
\section{\label{Chap4} Simulating the FMO Complex} 
\subsection{Energy transfer in FMO complex}\label{Chap4.energytransferFMO}
\begin{figure}
    \centering
    \includegraphics[width=0.8\linewidth]{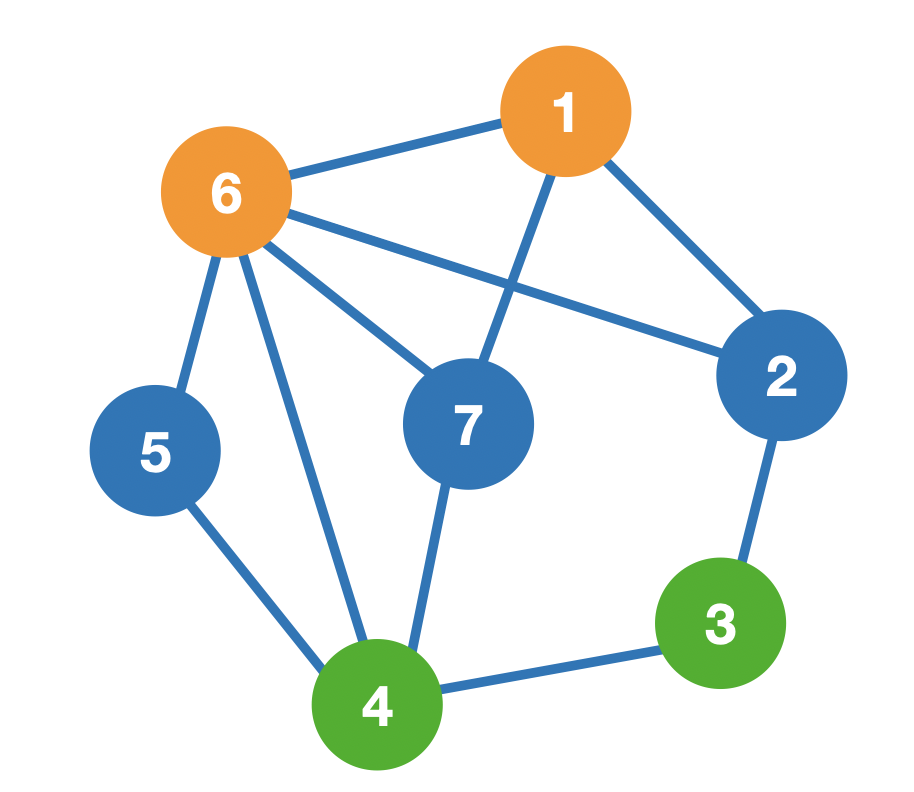}
    \caption{Representation of the 7 chromophore system of the FMO complex, the lines represent the dominant couplings between the sites. Site 6 and 1 represent the chromophores which receive energy from the light antenna, site 4 and 3 represent the sink which is connected to the reaction centre, and site 5, 7, and 2 represent the chromophores which form the intermediate path of the channel. Energy is transported from site 1 and 6 to site 3 and 4 via site 5, 7 and 2. The two dominant pathways are ($1\to2\to3$) and ($6\to(5,7)\to4\to3$).}
    \label{fig:fmo}
\end{figure}

FMO complex is the quantum transport channel for excitation energy transfer in green sulphur bacteria. The complex contains chromophores, which act as sites for excitons and are held by a protein scaffold at the right distances and orientations for efficient energy transfer. 
The sites show quantum coherence. The Hamiltonian\,\cite{mohseini} for the multi chromophoric system is given by,
\begin{equation}\label{eq:hamiltonian}
    H_c =\sum_{m=1}^{N_c} \varepsilon_m a_m^{\dagger}a_m + \sum_{n<m}^{N_c} V_{mn}(a_m^{\dagger}a_n + a_n^{\dagger}a_m).
\end{equation}
$N_c =$  7 is the number of chromophores in FMO complex. The $a_m^{\dagger}$ and $a_m$ are the creation and annihilation operators for an electron-hole pair (exciton) at chromophore m and $\varepsilon_m $ are the site energies. $V_{mn}$ are Coulomb couplings of the transition densities of the chromophores.  At any time  there is one exciton in the complex. Initial excitation occurs at site 1 or 6 and is transported to the sink at sites 3 and 4. The structure of the FMO complex\,\cite{fmo} is shown in Fig. \ref{fig:fmo}. Dominant couplings are represented by edges in Fig. \ref{fig:fmo} for which $V_{mn}$ is large. The  channel  is  subjected  to  noise  by  the  environment.
The phonon bath (protein scaffold) induces quantum jumps, decoherence and dephasing of excitons without changing the number of excitations. The  phonon coupling Hamiltonian is,
\begin{equation*}\label{eq:phonon}
    H_p=\sum_{m,n}^{N_c} q_{mn}^pa_m^{\dagger}a_n.
\end{equation*}
Damping of the excitation due to the radiation field is given by the Hamiltonian,
\begin{equation*}\label{eq:damping}
    H_r=\sum_{m}^{N_c} q_{m}^r(a_m^{\dagger}+a_m).
\end{equation*}
Lamb shifts due to phonon and photon bath coupling contribute negligibly to the dynamics\,\cite{adolph}, so are excluded from the equations. The Lindblad master equation in the Born-Markov and secular approximations is given by,
\begin{equation}\label{eq:master}
    \frac{\partial\rho(t)}{\partial t} = -\frac{i}{\hbar}[H_c,\rho(t)] + L_p(\rho(t)) + L_r(\rho(t)).
\end{equation}
The respective Lindblad superoperators $L_p$ and $L_r$ are given by,
\begin{equation}\label{eq:lindblad}
\begin{split}
    L_k(\rho(t))= &\sum_\omega\Gamma^k(\omega)\sum_{m,n}\Big[A_m^k(\omega)\rho A_n^{k\dagger}(\omega)\\
    &-\frac{1}{2}A_m^k(\omega)A_n^{k\dagger}(\omega)\rho - \frac{1}{2}\rho A_m^k(\omega)A_n^{k\dagger}(\omega)\Big] 
\end{split}
\end{equation}
Here ($k=p,r$), $A_m^p(\omega) =\sum_{\Omega-\Omega'=\omega}c_m^*(M_\Omega) * c_m(M_{\Omega'})|M_\Omega \rangle  \langle M_{\Omega'}|$, where $|M_\Omega \rangle$ is the exciton with frequency $\Omega$  and $|M_\Omega\rangle =\sum_mc_m(M_\Omega)|m\rangle$. The exciton states and their energies are the eigenvectors and eigenvalues, obtained by diagonalizing Hamiltonian $H_c$ (given in Eq.\eqref{eq:hamiltonian}). Excitons are delocalised over sites and the $A_m^p(\omega)$ represent delocalised exciton transport. The system Hamiltonian delocalises the excitons over sites due to coherence and phonon bath induces relaxation of these delocalised excitons. The jump operators constructed from exciton states capture the effect of coherence on {dissipative}  transport. Quantum coherence causes excitons to delocalise over different sites and the quantum jumps act between these delocalised exciton states rather than between different sites. The equation has both quantum and {dissipative} effects and captures their interplay which leads to greater transport efficiency. $\Gamma^p(\omega)$ are the rates for quantum jumps\,\cite{cho}given by,
\begin{equation}\label{eq:Gamma}
    \Gamma^p(\omega)=2\pi J(\omega)(1+n(\omega))
\end{equation} 
where $J(\omega)$ is Ohmic spectral density and $n(\omega)=1/[exp(\hbar\omega/kT)-1]$. The rate of quantum jumps increases with temperature. This can be understood in terms of spontaneous and induced relaxations caused by the environment. As the temperature increases, the probability of induced relaxation increases, which cause more quantum jumps and thus the phonon coupling is higher.\\
Damping is of the order of 1 $ns^{-1}$\,\cite{damping} and the transfer time for excitation across the complex is $\sim$ 4 ps\,\cite{adolph}. Since the damping contribution is negligible for the duration of the quantum walk, it is neglected in the analysis. Overall, the dynamics are given by,
\begin{equation}\label{eq:masterfmo}
    \frac{\partial\rho(t)}{\partial t} \approx -\frac{i}{\hbar}[H_c,\rho(t)] + L_p(\rho(t)).
\end{equation}
The above is similar to the ENAQT quantum walk described in Eq.(\ref{eq:masterENAQT}). Thus, the dynamics in the FMO complex are described by the master equation for ENAQT. We can treat the multichromophoric system in the FMO complex as an open quantum system. The fluctuations of correlated protein environment form a quantum bath which enhances the energy transport in the FMO complex\,\cite{nesterov2015, kurt2020}. Due to coherence, excitons delocalize over multiple chromophores. This facilitates the quantum jumps between excitons. Quantum walks can give an exponential speedup over classical walks due to interference which speeds up the energy transfer\,\cite{childs2003exponential}. In the next section, we give the theoretical framework for digital quantum simulation of the dynamics of the FMO complex using the general solution, Eq.(\ref{eq:case2operatoreq}), of the master equation developed in Sec.\ref{Chap3.analytical}.
\vspace{1cm}
\subsection{Theoretical model for simulating FMO complex}\label{Chap4.model}
The quantum jumps, analogous to Lindblad operators, Eq.(\ref{eq:lindblad}) show delocalised exciton transport and can be represented as follows.
Say, the system is in the exciton state $|M\rangle$. If the probability of quantum jump to state $|N\rangle$ in time $\sqrt{\Delta t}$ is $\gamma_{M\to N}$ (obtained from $\Gamma^k(\omega)$), for any $|N\rangle \neq |M\rangle$, then the Kraus operators for quantum jumps from $|M\rangle$ are given by:
\begin{equation}\label{eq:fmooperators}
\begin{split}
    M_{MN}&=\sqrt{\gamma_{M \to N}}|N\rangle \langle M| \quad \textrm{for all } N \neq M\\
    M_{MM}&= \sqrt{1-\sum_{N\neq M} \gamma_{M\to N}} |M\rangle \langle M|.
\end{split}
\end{equation}
Similar operators can be given for jumps from all such $|M\rangle$. Using the general model,  Eq.(\ref{case2operators}), and (\ref{eq:case2operatoreq}), the dynamical equation for the FMO complex is given by ($U=exp(-\frac{iH_c\Delta t}{\hbar})$),
\begin{equation}\label{eq:fmo}
\begin{split}
    &\rho(t +\Delta t)=\sum_M\Big[ M_{MM} U\rho(t) U^\dagger M_{MM}^\dagger\\
    &+\sum_{N  \neq M} \Big(M_{MN}\rho(t) M_{MN}^\dagger + M_{MM} U\rho(t) U^\dagger M_{NN}^\dagger\Big)\Big].
\end{split}
\end{equation}
As proven in Sec.\ref{Chap3.proof}, this equation is the solution for the  Lindblad equation for ENAQT in FMO complex. It should be noted that the last term in this equation is the decoherence term and the first and second terms are due to the quantum jumps. The evolution is trace preserving ($\sum_{i,j}M_{ij}^\dagger M_{ij}=\mathbb{1}$). This is the theoretical model for digital quantum simulation of the FMO complex.\bigskip
\begin{figure}[H]
    \centering
    \includegraphics[width=\linewidth]{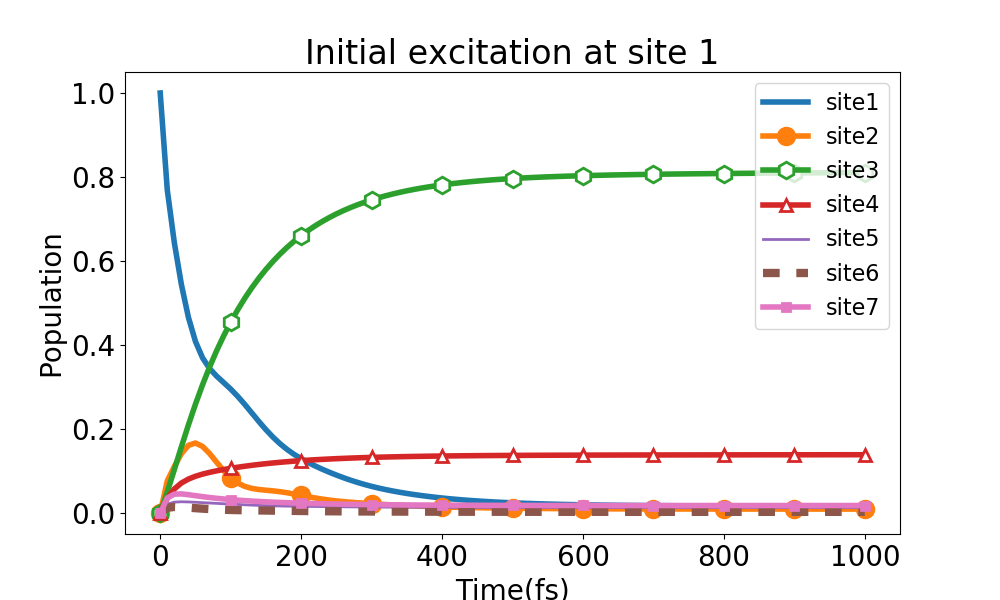} 
    \caption{Time evolution of the population at each site in the FMO complex calculated using Eq.\eqref{eq:fmo}. Initial Excitation at site 1, efficiency achieved=98$\%$.}
    \label{fig:1a}
    \centering
    \includegraphics[width=\linewidth]{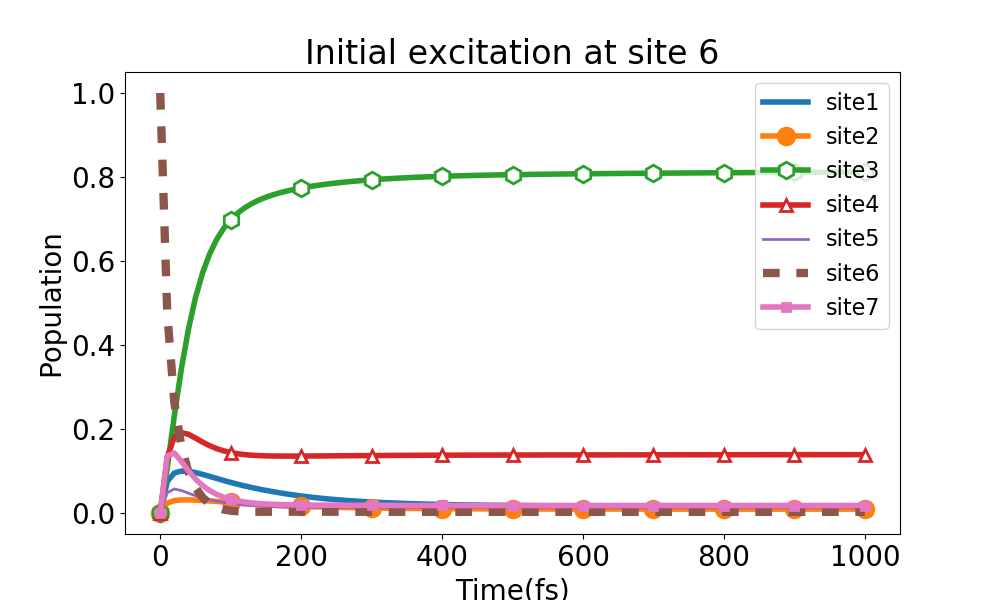}
    \caption{Time evolution of the population at each site in the FMO complex calculated using Eq.\eqref{eq:fmo}. Initial Excitation at site 6, efficiency achieved=98.3$\%$.}
    \label{fig:1b}
\end{figure}

\subsection{Numerical simulation}\label{Chap4.simulation}
Equations (\ref{eq:fmo}) and (\ref{eq:fmo_chi}) represent evolution of the density matrix in the FMO complex. To verify the effectiveness of these models, we simulate them numerically and present the results. Evolution of population densities at different sites is calculated using Eq.(\ref{eq:fmo}).The step size ($\Delta t$) is chosen to be 10 fs, as observed coherence time is $\sim 300$ fs and exciton relaxation time is $\sim 70$ fs\,\cite{engel}. For the numerical simulation, the value of the system Hamiltonian is taken from\,\cite{cho}. The quantum jump rates between excitons are calculated from relaxation data presented in\,\cite{7}, using theoretical analysis given in\,\cite{adolph,cho}.
\subsubsection{Result: High efficiency of energy transfer}\label{Chap4.highefficiency}
In Figures\,\ref{fig:1a}, \ref{fig:1b} we show the evolution of population densities when the initial excitation is at site 1 (Fig.\,\ref{fig:1a}) and site 6 (Fig.\,\ref{fig:1b}). The efficiency attained (sum of population on site 3 and site 4  at t = $4$ ps) is $\sim 98 $ percent. It is in agreement with theoretical evidence presented in \,\cite{mohseini}, by using master equation model, Eq.(\ref{eq:master}).  It can also be seen from the graphs that transfer happens faster if the initial excitation is at site 6, as observed experimentally in \,\cite{adolph}.

\begin{figure}[H]
    \centering
    \includegraphics[width=\linewidth]{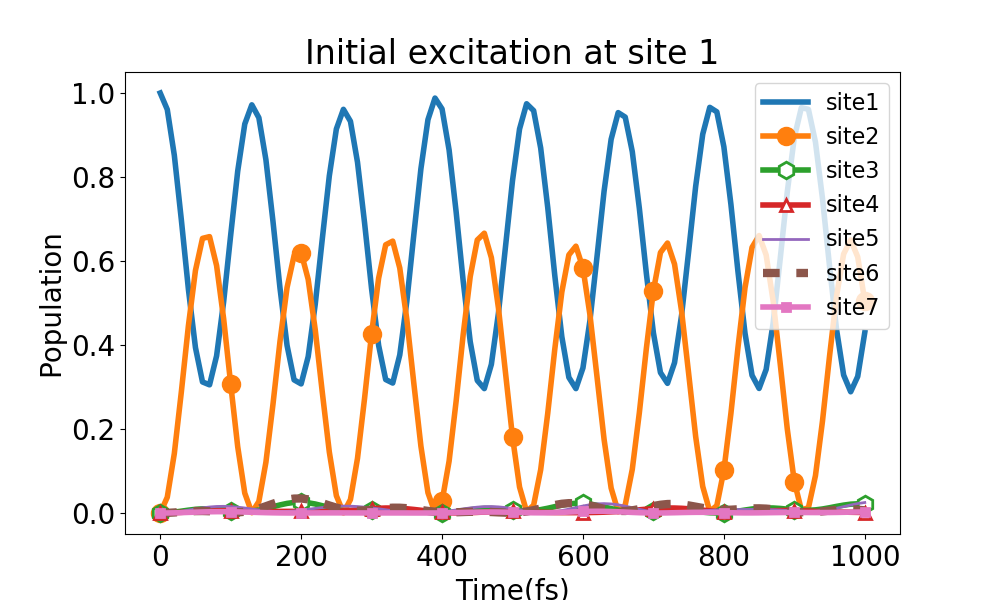}
    \caption{Time evolution of the population at each site in the FMO complex under purely coherent effects. Initial Excitation at site 1.}
    \label{fig:localization1}
\end{figure}
\begin{figure}[H]
    \centering
    \includegraphics[width=\linewidth]{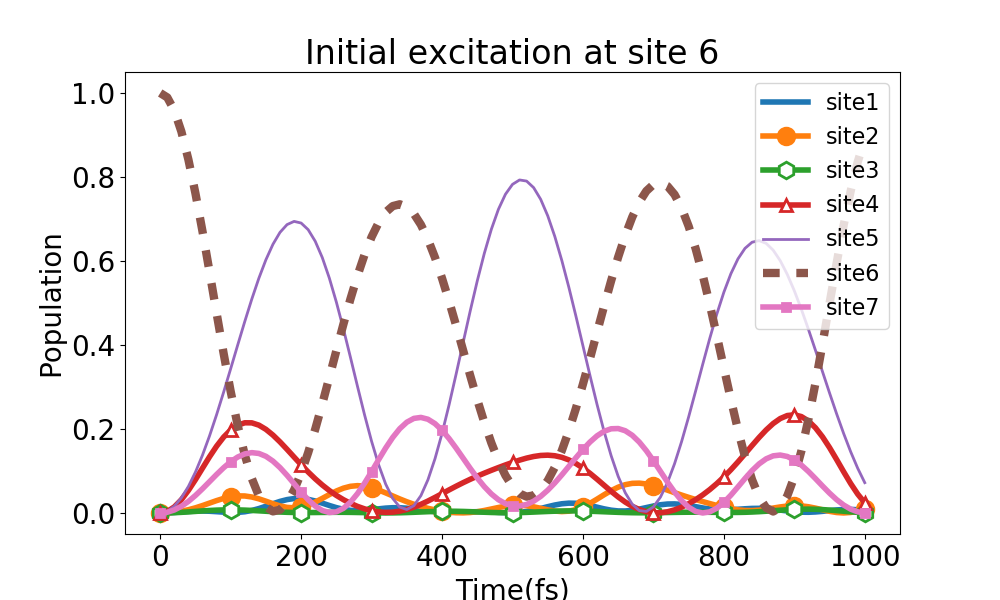}
    \caption{Time evolution of the population at each site in the FMO complex under purely coherent effects. Initial Excitation at site 6.}
    \label{fig:localization6}
\end{figure}

\subsubsection{Directionality in the quantum walk}
In the FMO complex, directionality of energy transport is given by quantum jumps which cause the transfer of exciton from initial sites towards the sink. To show this, we simulate the evolution in absence of quantum jumps (Fig.\,\ref{fig:localization1}, Fig.\,\ref{fig:localization6}). As we can see, the population just oscillates between initial strongly coupled sites, that is it stays at the initial exciton. This is caused by Anderson localization due to disorder in the Hamiltonian of the FMO complex. Quantum jumps help overcome the trapping of exciton due to Anderson localization by transferring the population between different excitons. The probabilities of quantum jumps are greater in the direction from initial site towards the final site, thus giving directionality to the walk. Compared to these figures, the simulations with quantum jumps(Fig.\,\ref{fig:1a} and Fig.\,\ref{fig:1b}) show exciton transport. The role of noise in ENAQT is also illustrated in\,\cite{chin2010noise}.
\subsubsection{Result: Dependence on environment temperature}\label{Chap4.temp}
\begin{figure}[H]
    \centering
    \includegraphics[width=\linewidth]{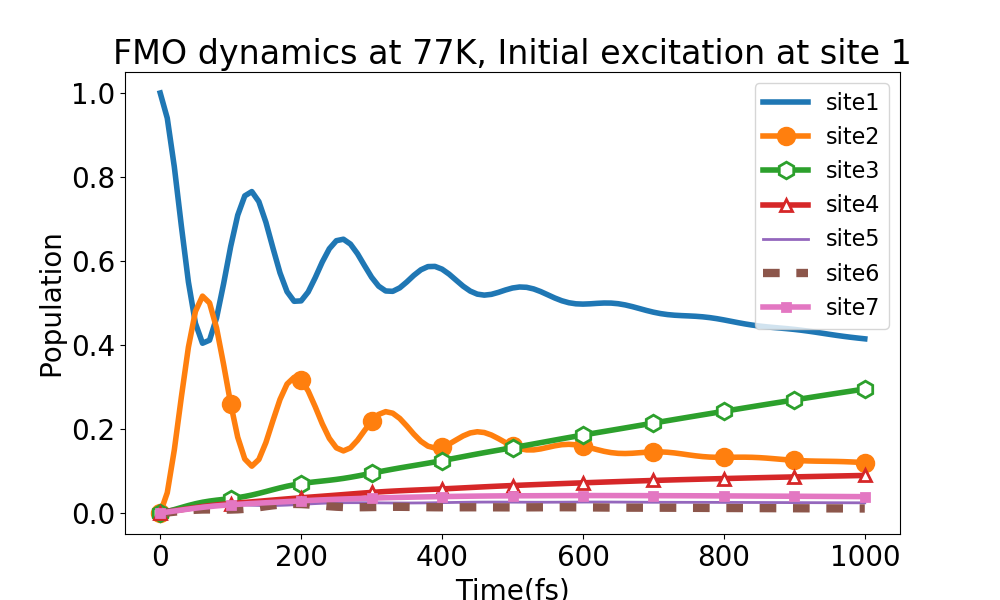} 
    \caption{Time evolution of the population at each site in the FMO complex calculated using Eq.\eqref{eq:fmo_chi}. Initial Excitation at site 1, $\chi=0.06$.}
    \label{fig:2a}
\end{figure}

\begin{figure}[H]
    \centering
    \includegraphics[width=\linewidth]{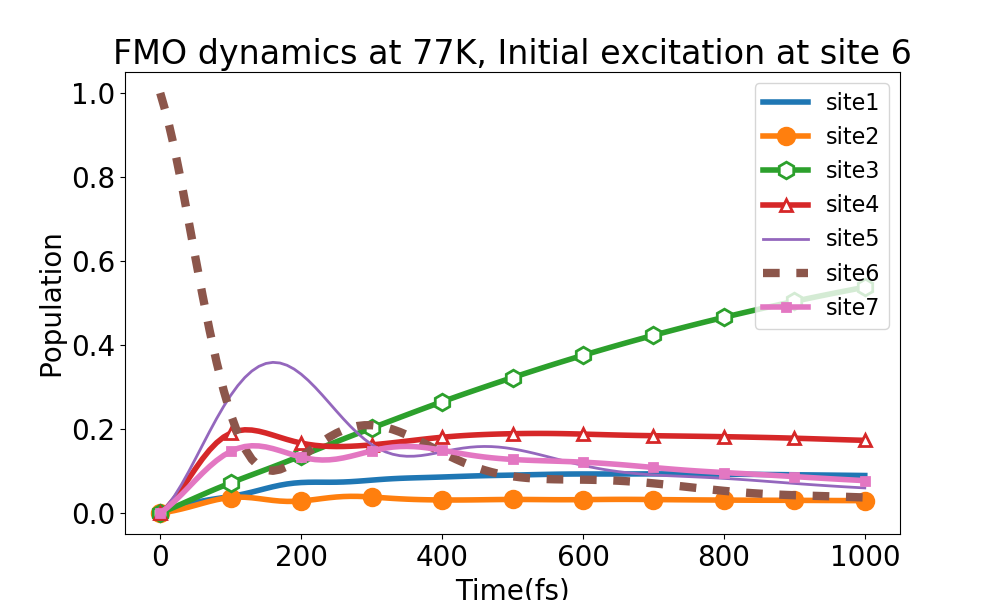}
    \caption{Time evolution of the population at each site in the FMO complex calculated using Eq.\eqref{eq:fmo_chi}. Initial Excitation at site 6, $\chi=0.06$.}
    \label{fig:2b}
\end{figure}

The dynamics of the FMO complex are temperature dependent. The phonon couplings vary with temperature and the coupling strength can be seen to increase as the temperature increases, Eq.(\ref{eq:Gamma}). The dynamics in the FMO  complex have been observed upto ambient temperatures. At lower temperatures, the coupling strength is a fraction of the maximum coupling constants. This can be digitally simulated by the dynamical equation with tunable bath coupling, Eq.\,(\ref{eq:fmo_chi}).  In Figures \ref{fig:2a} and \ref{fig:2b} we show the results of simulations with different phonon couplings, Eq.(\ref{eq:fmo_chi}). For $\chi=0.06$ the evolution of population densities at different sites are shown when the initial excitation at site 1 (Fig.\ref{fig:2a}) and site 6 (Fig.\ref{fig:2b}). 
In Fig.\ref{fig:2a} it can be seen that there is an initial oscillation of population between site 1 and 2. This is due to the high coupling between these two sites which causes the exciton over site 1 to delocalise to site 2. Slowly this oscillation dies, as quantum jumps cause the population to move towards the sink at sites 3 and 4, whose population starts rising. There is some population at site 7 as well, which serves as the connecting link between different sites (as can be seen in Fig.\ref{fig:fmo}). As compared to Fig.\ref{fig:1a}, there is an enhanced effect of quantum dynamics (coherence between site 1 and 2) visible in Fig.\ref{fig:2a}. This is expected since the strength of the bath coupling is lower for $\chi=0.06$, leading to suppression of environment induced quantum jumps. Similar dynamics can be seen in Fig.\ref{fig:2b} where there is an initial oscillation of population between site 6 and site 5. This population is slowly transported towards the sink at site 3 and 4.

Figures \ref{fig:2a} and \ref{fig:2b} represent the dynamics in the FMO complex at 77K. These results can be matched with the dynamics obtained in previous theoretical study\,\cite{ishizaki}. We can see that the results of calculations done using the discrete time evolution model of Eqs.(\ref{eq:fmo})  and (\ref{eq:fmo_chi}) match with theoretical\,\cite{ishizaki,mohseini} and experimental studies\,\cite{adolph} done on the FMO complex.

\subsection{\label{secImplementation}Quantum circuit for implementation of the model}
The usual method for implementation of the FMO complex is to treat each site (position basis) on a separate qubit, whose excitation from the ground state $\ket{0}$ to excited state $\ket{1}$ using creation operator ($a^\dagger$) depicts population at that site. This requires $n$ qubits with and Hilbert space of the dimension $2^{n}$ for implementing a quantum system with n sites.  However, the Hamiltonian of the system, Eq.\eqref{eq:hamiltonian} has dimension $n$. Mapping the Hamiltonian to the expanded basis for implementation requires complex calculations and often multiple techniques have to be employed for simulation. Additionally, for implementing the non unitary part of evolution, each qubit of the system uses an ancilla qubit to capture dephasing of the qubit. This approach gives an $O(n)$ space complexity of implementation. 

However, simulation of the framework presented in this paper can be done on a smaller qubit space. {The FMO complex represents a seven site system and the state of exciton on the sites can be represented by a Hilbert space with $dim=7$. Minimum number of qubits required for simulating this is $\lceil\log 7\rceil=3$. The seven exciton states can be represented as, \\
\begin{equation}\label{eq:excitonstates}
    \begin{split}
        &E_1=\ket{001}\qquad E_5=\ket{101}\\
        &E_2=\ket{010}\qquad E_6=\ket{110}\\
        &E_3=\ket{011}\qquad E_7=\ket{111}.\\
        &E_4=\ket{100}
    \end{split}
\end{equation}
We can implement one quantum jump, say from exciton state $E_i$ to $E_j$ as
\begin{equation}
    \rho\to \gamma_{i\to j}\ket{j}\bra{i}\rho\ket{i}\bra{j},
\end{equation}
where $\ket{i}$ represents exciton $E_i$ and $\gamma_{i\to j}$ is probability of quantum jump from state $E_i$ to $E_j$. \\
Overall, the dynamics can be represented as,
\begin{equation}
\begin{split}
    \ket{i}\ket{0}_B\to\sqrt{1-\gamma_{i\to j}}\ket{i}\ket{0}_B&\\
    +\sqrt{\gamma_{i\to j}}\ket{j}\ket{1}_B&.
\end{split}
\end{equation}
We can modify it to include another bath qubit as,
\begin{equation}
\begin{split}
    \ket{i}\ket{0}_{B_1}\ket{0}_{B_2}\to\sqrt{1-\gamma_{i\to j}}\ket{i}\ket{0}_{B_1}\ket{0}_{B_2}&\\
    +\sqrt{\gamma_{i\to j}}\ket{j}\ket{1}_{B_1}\ket{1}_{B_2}&.
\end{split}
\end{equation}
This can also be written as,
\begin{equation}
\begin{split}
    \ket{0}_{B_1}\ket{i}\ket{0}_{B_2}\to\sqrt{1-\gamma_{i\to j}}\ket{0}_{B_1}\ket{i}\ket{0}_{B_2}&\\
    +\sqrt{\gamma_{i\to j}}\ket{1}_{B_1}\ket{j}\ket{1}_{B_2}&.
\end{split}
\end{equation}
The above is for one quantum jump. We can trace out the second bath qubit and obtain the Kraus operators,
\begin{equation}\label{eq:Krausij}
\begin{split}
    M_0^{ij}=(\sqrt{1-\gamma_{i\to j}})\ket{0,i}\bra{0,i} + \ket{1,i}\bra{1,i} &\\
    +I_{B_1}\otimes\sum_{m\neq i} \ket{m}\bra{m} &
\end{split}
\end{equation}
~
\begin{equation}
    M_1^{ij}=\sqrt{\gamma_{i\to j}}\ket{1,j}\bra{0,i}
\end{equation}
where $\ket{p,q}=\ket{p}_{B_1}\ket{q}$ for $p\in \{0,1\}$ and $q\in [1,7]$. The above Kraus operators for one quantum jump leave all states $\ket{m}\neq\ket{i}$ and the state $\ket{1,i}$ unchanged and implement the quantum jump from $\ket{0,i}\to\ket{1,j}$.}
\begin{widetext}
{Each step of the evolution consists of multiple quantum jumps, which can be represented as,
\begin{equation}
    \begin{split}
        &\text{$1^{st}$ jump (from $\ket{1}\to\ket{2}$)}: \\
        &\qquad M_0=(\sqrt{1-\gamma_{1\to2}})\ket{0,1}\bra{0,1}+ \ket{1,1}\bra{1,1}+I_{B_1}\otimes\sum_{m\neq 1} \ket{m}\bra{m} \qquad M_1=\sqrt{\gamma_{1\to2}}\ket{1,2}\bra{0,1}\\
        &\text{$2^{nd}$ jump (from $\ket{1}\to\ket{3}$):} \\
        &\qquad M_0=(\sqrt{1-\gamma_{1\to3}})\ket{0,1}\bra{0,1}+ \ket{1,1}\bra{1,1}+I_{B_1}\otimes\sum_{m\neq 1} \ket{m}\bra{m} \qquad M_1=\sqrt{\gamma_{1\to3}}\ket{1,3}\bra{0,1}\\
        &\text{$3^{rd}$ jump (from $\ket{1}\to\ket{4}$):} \\
        &\ldots\\
        &\ldots\\
        &\text{$7^{th}$ jump (from $\ket{2}\to\ket{1}$)}: \\
        &\qquad M_0=(\sqrt{1-\gamma_{2\to1}})\ket{0,2}\bra{0,2}+ \ket{1,2}\bra{1,2}+I_{B_1}\otimes\sum_{m\neq 2} \ket{m}\bra{m} \qquad M_1=\sqrt{\gamma_{2\to1}}\ket{1,1}\bra{0,2}\\
        &\text{$8^{th}$ jump (from $\ket{2}\to\ket{3}$):} \\
        &\qquad M_0=(\sqrt{1-\gamma_{2\to3}})\ket{0,2}\bra{0,2}+ \ket{1,2}\bra{1,2}+I_{B_1}\otimes\sum_{m\neq 2} \ket{m}\bra{m} \qquad M_1=\sqrt{\gamma_{2\to3}}\ket{1,3}\bra{0,2}\\
        &\ldots\\
        &\ldots\\
        &\ldots\\
        &\text{$42^{nd}$ jump (from $\ket{7}\to\ket{6}$):} \\
        &\qquad M_0=(\sqrt{1-\gamma_{7\to6}})\ket{0,7}\bra{0,7}+ \ket{1,7}\bra{1,7}+I_{B_1}\otimes\sum_{m\neq 7} \ket{m}\bra{m} \qquad M_1=\sqrt{\gamma_{7\to6}}\ket{1,6}\bra{0,7}.\\
    \end{split}
\end{equation}
}
\end{widetext}
{The following proof shows that sequentially implementing these jumps captures the Markovian evolution of the system under multiple quantum jumps. Consider any two quantum jumps $\ket{u}\to\ket{v}$ and $\ket{w}\to\ket{x}$. Sequential application of Kraus operators gives,\\
\begin{equation}
    \begin{split}
        \tilde{\rho}=M_0^{uv}(\rho(0)){M_0^{uv}}^\dagger + M_1^{uv}(\rho(0)){M_1^{uv}}^\dagger\\
        {\rho}(t)=M_0^{wx}(\tilde{\rho}){M_0^{wx}}^\dagger + M_1^{wx}(\tilde{\rho}){M_1^{wx}}^\dagger
    \end{split}
\end{equation}
where $M_{0/1}^{uv}$ are Kraus operators for jump from $\ket{u}\to\ket{v}$. Thus,
\begin{equation}
\begin{split}
    \rho\to &\quad M_0^{wx}M_0^{uv}(\rho){M_0^{uv}}^\dagger{M_0^{wx}}^\dagger 
    + M_0^{wx}M_1^{uv}(\rho){M_1^{uv}}^\dagger{M_0^{wx}}^\dagger\\
    &+M_1^{wx}M_0^{uv}(\rho){M_0^{uv}}^\dagger{M_1^{wx}}^\dagger 
    + M_1^{wx}M_1^{uv}(\rho){M_1^{uv}}^\dagger{M_1^{wx}}^\dagger.
\end{split}
\end{equation}
Now for general $\{u,v\}\neq\{w,x\}$ , using Eq.\,\eqref{eq:Krausij}
\begin{equation}
    \begin{split}
        M_0^{wx}M_0^{uv}=&(\sqrt{1-\gamma_{w\to x}})\ket{0,w}\bra{0,w}\\
        &+(\sqrt{1-\gamma_{u\to v}})\ket{0,u}\bra{0,u}\\
        &+\sum_{m\neq w,u} I_{B_1}\otimes\ket{m}\bra{m}\\
        M_0^{wx}M_1^{uv}=&(\sqrt{\gamma_{u\to v}})\ket{1,v}\bra{0,u}\\
        M_1^{wx}M_0^{uv}=&(\sqrt{\gamma_{w\to x}})\ket{1,x}\bra{0,w}\\
        M_1^{wx}M_1^{uv}=&0.
    \end{split}
\end{equation}
We can see that $M_0^{wx}M_1^{uv}$ and $M_1^{wx}M_0^{uv}$ represent the quantum jumps and $M_0^{wx}M_0^{uv}$ gives the population that remains for coherent evolution. This is exactly like the general model of ENAQT developed in the paper. The above operators are calculated for $\{u,v\}\neq\{w,x\}$. Similar calculations can be done when this condition does not hold. Thus, sequentially applying Kraus operators gives the Markovian evolution of the system, as described by ENAQT for 2 quantum jumps. The same argument can be extrapolated for multiple quantum jumps implemented one after the other. \begin{figure}[H]
    \centering
    \includegraphics[width=0.8\linewidth]{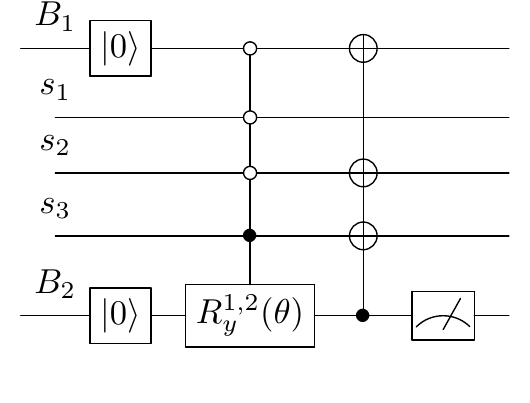}
    \caption{Quantum circuit for implementing one quantum jump from exciton $\ket{1}\to\ket{2}$ of the FMO complex. $s_1, s_2, s_3$ represent the system. Each exciton state is $\ket{s_1 s_2 s_3}$. For example, $\ket{1}=\ket{001}$. $B_1$ and $B_2$ are the bath qubits.}
    \label{fig:onejump}
\end{figure}
In the procedure described above, $\ket{0}_{B_1}$ is entangled with the population remaining after quantum jumps and $\ket{1}_{B_1}$ is entangled with the part of system capturing quantum jumps.\\
The coherent evolution due to Hamiltonian of the system can be implemented at the end of all quantum jumps using state $\ket{0}_{B_1}$ as,
\begin{equation}
    C=\ket{0}_{B_1}\bra{0}_{B_1}\otimes U + \ket{1}_{B_1}\bra{1}_{B_1}\otimes I
\end{equation}
where $C$ is the unitary operator that acts on $B_1$ and the system to evolve the coherent part of density matrix (entangled with $\ket{0}_{B_1}$) under unitary evolution $U=e^{-i\frac{H_c}{\hbar}t}$.\\ \\ 
$B_1$ is traced out at the end to complete one step of evolution with both coherent evolution and multiple quantum jumps.\\
\noindent The main advantage of using two bath qubits is that it helps us implement simultaneous quantum jumps sequentially. This greatly simplifies the simulation procedure. Overall, the second bath qubit is used to implement quantum jumps and the first bath qubit is used to implement coherent evolution. The first qubit is traced out after one complete step of evolution consisting of coherent evolution and multiple quantum jumps (which are implemented by tracing out the second bath qubit after each quantum jump and resetting it to $\ket{0}_{B_2}$).

\noindent The circuit for simulating one quantum jump is given in Fig.\ref{fig:onejump}. Here, an example is given for quantum jump from $\ket{1}\to\ket{2}$. $R_y^{1,2}(\theta)$ is the rotation about y-axis on the Bloch sphere such that $R_y^{1,2}(\theta)\ket{0}_{B_2}=\cos{\frac{\theta}{2}}\ket{0}_{B_2}+\sin{\frac{\theta}{2}}\ket{1}_{B_2}$ and  $\sin{\frac{\theta}{2}}=\sqrt{\gamma_{1\to 2}}$. The evolution of state under these operations is,
\begin{equation}\label{eq:quantumjump12}
    \begin{split}
        &\text{First gate}:\\
        &\qquad\ket{0}_{B_1}\ket{001}\ket{0}_{B_2}\to\sqrt{1-\gamma_{1\to 2}}\ket{0}_{B_1}\ket{001}\ket{0}_{B_2}\\
        &\qquad~\qquad~\qquad\qquad\qquad+\sqrt{\gamma_{1\to 2}}\ket{0}_{B_1}\ket{001}\ket{1}_{B_2}.\\
        &\text{Second gate}:\\
        &\qquad\sqrt{\gamma_{1\to 2}}\ket{0}_{B_1}\ket{001}\ket{1}_{B_2}\to\sqrt{\gamma_{1\to 2}}\ket{1}_{B_1}\ket{010}\ket{1}_{B_2}\\
        &\text{which can also be written as (using Eq.\,\eqref{eq:excitonstates})}\\
        &\qquad\sqrt{\gamma_{1\to 2}}\ket{0}_{B_1}\ket{1}\ket{1}_{B_2}\to\sqrt{\gamma_{1\to 2}}\ket{1}_{B_1}\ket{2}\ket{1}_{B_2}.
    \end{split}
\end{equation}
\begin{figure}[H]
    \centering
    \includegraphics[width=0.5\linewidth]{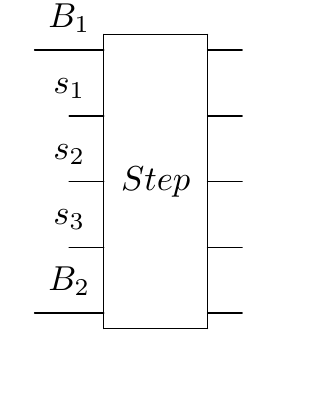}
    \caption{Each step of evolution, denoted by a  "Step" operator consists of multiple quantum jumps. Thus, the "Step" operator is equivalent to the circuit given in Fig.\ref{fig:onestep}.}
    \label{fig:stepoperator}
\end{figure}

\begin{figure*}[ht]
    \centering
    \includegraphics[width=\textwidth]{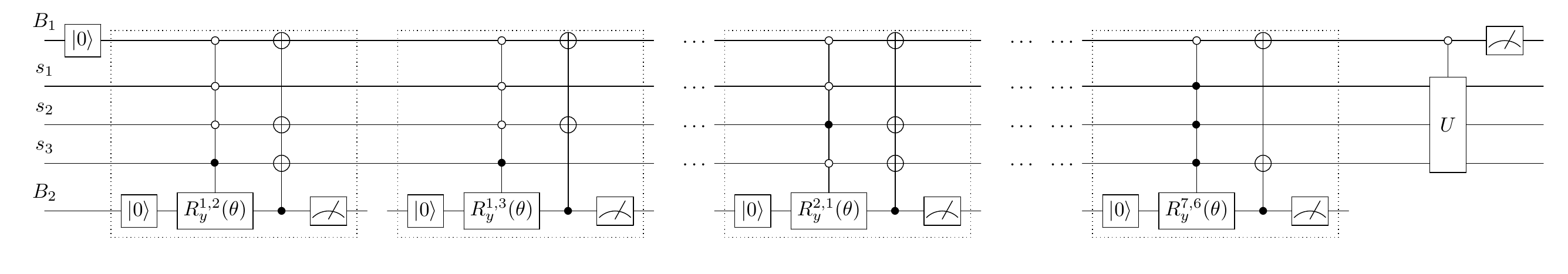}
    \caption{Quantum circuit for implementing multiple quantum jumps from exciton $\ket{i}\to\ket{j}$ for $i,j\in[1,7]$. The whole circuit makes one step of evolution.}
    \label{fig:onestep}
\end{figure*}
\noindent The circuit for simulating one step (Fig.\ref{fig:stepoperator}) of evolution of the FMO complex is given in Fig.\ref{fig:onestep}, constructed by composing quantum jumps sequentially. \\

Fig.\ref{fig:FMO} shows the complete circuit diagram for multiple steps. Apart from the quantum jumps and coherent evolution, there is an initial gate (D) which transforms the basis from position to excitons which is inverted at the end before measurement of population at different sites. This is done to aid our calculations for quantum jumps which are done in the exciton basis. The gaps between steps in $B_1$ qubit are to indicate that it is traced out at the end of each step and reset to $\ket{0}_{B_1}$ at the beginning of the next step
\begin{figure}[H]
    \centering
    \includegraphics[width=\linewidth]{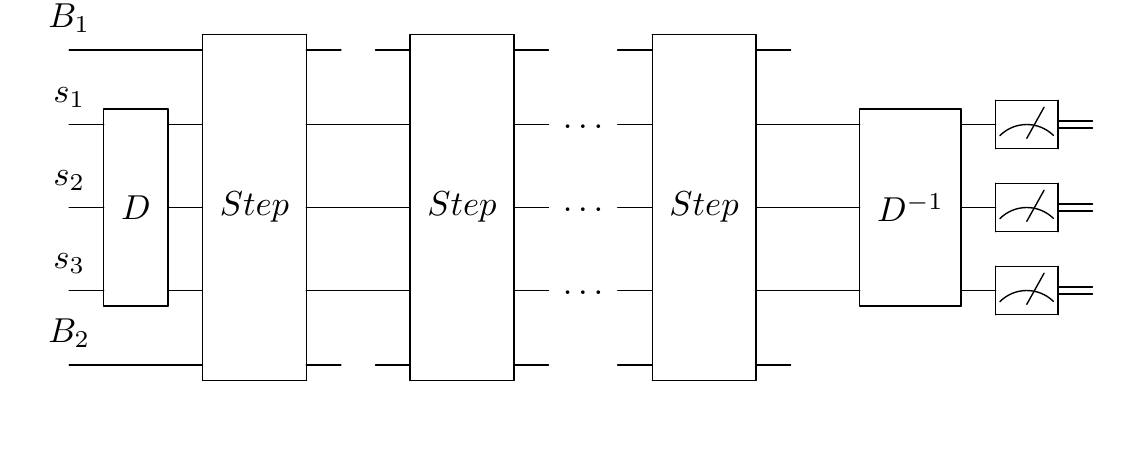}
    \caption{Complete evolution of the FMO complex can be implemented using multiple "Step" operators. The state is initiallized in qubits $s_1, s_2, s_3$. Operator "D" transforms the state from position basis to exciton basis. After simulation, the operator "$D^{-1}$" rotates the final state back to position basis. Final state can be measured by measuring the system qubits. }
    \label{fig:FMO}
\end{figure}
\begin{figure}[H]
    \centering
    \includegraphics[width=0.75\linewidth]{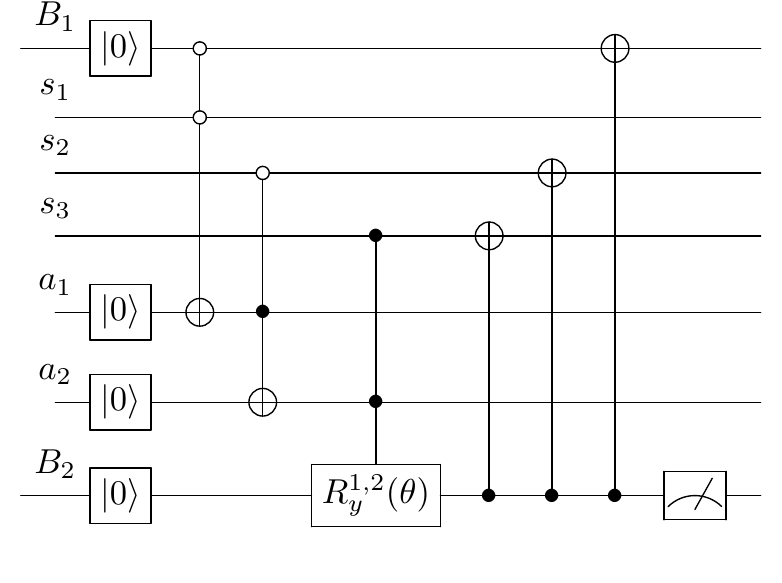}
    \caption{Operators for a single quantum jump from $\ket{1}\to\ket{2}$, given in Fg.\,\ref{fig:onejump} are decomposed into elementary operators for analyzing the complexity. $a_1, a_2$ represent the ancilla introduced in the circuit to implement multi qubit controls using two qubit control operators.}
    \label{fig:decomposition}
\end{figure}

\noindent {\it Complexity analysis}:\\

\noindent {\it Number of gates:} To analyze the complexity in terms of number of gates, we need a to decompose multi qubit gates to elementary gates. Thus, for single quantum jump, Fig.\ref{fig:onejump} can also be drawn as Fig.\ref{fig:decomposition}. We can see that we require $\log n$ gates for the first gate (Eq.\ref{eq:quantumjump12}) and $\log n$ gates for the second gate (Eq.\ref{eq:quantumjump12}). Similar decomposition can be done for all quantum jumps shown in Fig.\ref{fig:onestep}. \\
\\
There are $n(n-1)$ quantum jumps, each of which requires $2\log n$ gates. This gives $O(n^2\log n)$ gates for implementing quantum jumps. There is one gate for implementing the coherent evolution. Decomposition of unitary evolution in terms of elementary gates will take $O(n)$ gates (decomposition of arbitrary unitary gate over $q$ qubits takes $O(exp(q))$ gates, and we have $q=\log n$ qubits). Thus the gate complexity for implementing one step of evolution is $O(n^2\log n)$. The number of gates is also smaller than needed in traditional Stinespring dilation of quantum channels which often need $O(n^6)$ gates\,\cite{6}. The total complexity for complete evolution for time=$T$ is $O(\frac{T}{\Delta t}n^2\log n)$, where one time step=$\Delta t$. \\
\\ \\
{\it Number of qubits:}  We have used $\log n$ qubits for simulating the system and 2 bath qubits. Additionally, we use $\log n$ ancillas for simulating multi qubit control gates. Thus we need $O(\log n)$ qubits for simulating the FMO complex. The extra qubits can be avoided by using Suzuki-Lie Trotter decomposition of the gates. \\ \\ 
\noindent We have provided a model for digital quantum simulation of energy transfer in open quantum systems (the approach can be applied to systems other than FMO complex). The alternate to such simulations is to use a Stinespring dilation for getting a unitary quantum evolution of system and bath which can simultaneously implement the quantum jumps. An approximate decomposition of the unitary evolution as elementary one and two qubit gates is then obtained using techniques like Suzuki-Lie Trotter decomposition. Or each site is treated on a separate qubit along with an ancillary qubit and the effective evolution is mapped on a larger Hilbert space of $7$ qubits ($dim=2^7$) + $7$ ancilla qubits. However, our framework only uses $\lceil\log 7\rceil=3$ qubits thus giving a log-reduction in space. There is also a huge reduction is space of ancillary qubits from $O(n)$ or $O(\log n)$ to just 2 qubits. This reduction in complexity from $poly(n)$ to $O(1)$ gives the absolute minimum space complexity that can implement the bath. 
\\ }


Thus, the discrete time evolution equation, together with a rather straightforward way of varying the coupling strength of environment provides an optimal method to digitally simulate environment assisted transport in open quantum systems.

\section{\label{sec5}Concluding remarks}
We have developed a theoretical framework for digital quantum simulation of the environment assisted quantum transport(ENAQT) in open quantum systems in the discrete-time density matrix evolution formalism. We modelled ENAQT into an open quantum systems and developed a methodology to to solve the Lindblad master equation for ENAQT.  We have obtained evolution operators, Eq.(\ref{case2operators})  to capture the interplay of quantum coherence and quantum jumps. Using the obtained evolution equation, Eq.(\ref{eq:case2operatoreq}) and the operators, we have proved its equivalence to the Lindblad master equation. We have shown that our approach is an improvement over the conventional numerical solution in which the quantum and {dissipative} effects are simulated separately. We applied the general solution of ENAQT master equation to simulate the FMO complex using delocalised exciton transport. Then, we generalised it to capture temperature dependence as variable strength of phonon coupling. We have demonstrated that calculations done on our model are in good agreement with experimental and theoretical evidence. Our model can be used to simulate the FMO complex dynamics for different temperatures and conditions. {We presented the quantum circuit model for it's implementation that gives a log reduction in complexity. }

Open quantum system are generally described using master equations for which solutions are otherwise not known. The solution of the master equation we developed can have great scope for applications to digital quantum simulation of ENAQT. The model can be used to understand noise assisted transport (NAT) in quantum simulators. It could help in developing systems with tunable level of noise. It can also be applied to study artificial photovoltaic quantum systems and quantum communication channels to achieve desired efficiency and other properties. The methodology developed to solve the Lindblad equation for ENAQT can be valuable in developing solution of master equation for other processes. It could be used to model multiple noise effects in complex systems with diverse sources of environmental interactions. Modelling effects such as depolarization and damping in open quantum systems, can be subjects for further research work. This can help in further developing the theory of digital quantum simulations. Another interesting line of work can be exploring the experimental implications of our work on developing quantum simulators. The framework presented in this paper and it's associated log-space reduction in complexity can be especially useful for simulating large scale systems. \\



\begin{acknowledgments}
CMC would like to thank Department of Science and Technology,
Government of India for the Ramanujan Fellowship grant No.:SB/S2/RJN-192/2014.
We also acknowledge the support from Interdisciplinary Cyber Physical
Systems (ICPS) programme of the Department of Science and Technology,
India, Grant No.:DST/ICPS/QuST/Theme-1/2019/1 and  US Army ITC-PAC contract no.  FA520919PA139. 
\end{acknowledgments}




\end{document}